\documentclass[useAMS,usenatbib,referee]{mn2e}

\usepackage{epsfig}

\title[Doppler imaging of LO Peg in 2003]{Doppler imaging of the young late-type star LO Pegasi 
(BD+22$^{\circ}$4409) in September 2003\thanks{Based on observations made with the
 Italian {\it Telescopio Nazionale Galileo} operated 
on the island of La Palma by the {\it Centro Galileo Galilei} of INAF ({\it Istituto Nazionale di Astrofisica}) 
at the Spanish {\it Observatorio del Roque del los Muchachos} of the {\it Instituto de Astrof\'{\i}sica de Canarias.}}}
\author[N.~Piluso et al.]{N.~Piluso$^{1,2}$\thanks{E-mails:
nicolo.piluso@oact.inaf.it (NP); nuccio.lanza@oact.inaf.it (AFL); isabella.pagano@oact.inaf.it (IP); alessandro.lanzafame@oact.inaf.it (ACL); donati@ast.obs-mip.fr (JFD).}, A.~F.~Lanza$^{2}$, I.~Pagano$^{2}$, A.~C.~Lanzafame$^{1}$, 
J.-F.~Donati$^{3}$\\
$^{1}$Sezione Astrofisica, Dipartimento di Fisica e Astronomia, Universit\`a degli Studi di Catania, Via S. Sofia, 78, 95123, Catania, Italy\\
$^{2}$INAF-Osservatorio Astrofisico di Catania, Via S. Sofia, 78, 95123, Catania, Italy \\
$^{3}$LATT - CNRS / Universite de Toulouse, 14 avenue E. Belin, F-31400 Toulouse, France
}
\begin{document}

\date{Accepted 2008 February 26; Received 2008 February 22; in original form 2008 January 11}

\pagerange{\pageref{firstpage}--\pageref{lastpage}} \pubyear{2007}

\maketitle

\label{firstpage}

\begin{abstract}
A Doppler image of the ZAMS late-type rapidly rotating star LO Pegasi, based on spectra acquired between 12 and 15 September 2003, is presented. The Least Square Deconvolution technique is applied to enhance the signal-to-noise ratio
of the mean rotational broadened line profiles extracted from the observed spectra. In the present application,
a unbroadened spectrum is used as a reference, instead of a simple line list, to improve the deconvolution technique
applied to extract the mean profiles. The reconstructed image is similar to those previously obtained from observations taken in 1993 and
1998, and shows that LO~Peg photospheric activity is dominated by high-latitude spots with a non-uniform polar cap. The
latter seems to be a persistent feature as it has been observed since 1993 with little modifications. 
Small spots, observed between $\sim 10^{\circ}$ and $\sim 60^{\circ}$ of latitude, appears to be different with respect to those present in the 1993 and 1998 maps.     
\end{abstract}


\section{Introduction}

Late-type rapidly rotating stars show phenomena characteristic of solar-like magnetic activity, such as 
photospheric spots, chromospheric line emissions, coronal X-ray and radio emissions, as well as flaring activity.
Their study has become important  to improve our understanding of the 
effects of magnetic fields in stellar atmospheres, to provide further constraints on dynamo models of magnetic field
generation, and to understand magnetic braking of stellar rotation.   

LO~Peg (BD+22$^{\circ}$4409) is a chromospherically active star of spectral class K5V-K7V,
detected during the ROSAT WFC EUV all-sky survey 
and the Extreme Ultraviolet Explorer (EUVE) survey.  The first study by \citet{Jeffersetal94} 
showed a high lithium abundance ([Li/H]=$1.30 \pm 0.25$) and a H$\alpha$ line in emission. 
They concluded that the star is young (with an age of approximately 30 Myr) and a likely member of the Local Association.

LO~Peg is an interesting target to study stellar activity because it is among the latest spectral type dwarfs
which have been Doppler imaged. However, only a few studies have specifically addressed the imaging of its surface
\citep[i.e., ][]{Listeretal99,Barnesetal05}.
 
For  stars as faint as LO~Peg ($m_{\rm V} = 9.2$), the achievement of a sufficient 
signal-to-noise  ratio for Doppler imaging represents the most serious problem  
because the exposure time must be kept within a few tens of minutes to avoid the smearing of the reconstructed map
due to fast stellar rotation. 
Therefore, we apply the method of LSD (Least Square Deconvolution) to increase the signal-to-noise ratio of LO~Peg spectra
\citep{Donatietal97}.
This method  makes use of a large number (up to $\sim 1500-2000$) of photospheric lines contained 
in an echelle spectrum to extract a mean
line profile with a signal-to-noise ratio sufficiently high to apply Doppler Imaging techniques. In such a way,
 it is possible to broaden
 the sample of Doppler imaging candidates by making accessible
 a greater volume of space including not only faint field objects, such as
LO~Peg, but also lower main-sequence stars in nearby open clusters. 

Our specific implementation of the LSD approach is described in Sect.~\ref{LSDdescr} 
and introduces some improvements with respect to the
original procedure by \citet{DonatiCameron97} { along the lines of \citet{Barnes05}}.

The first Doppler images of LO~Peg were obtained by \citet{Listeretal99} based on observations taken in August 1993.
They found two main regions of spot coverage:
a high latitude spot or polar crown with a starspot concentration towards longitude 70$^{\circ}$, 
and a low-latitude belt centred at a latitude of $25^{\circ}\pm 10^{\circ}$, with very little spot 
coverage in the mid-latitudes in between. Later images obtained by \citet{Barnesetal05}, based on data of July 1998, 
showed again a polar spot with a somewhat weaker longitude non-uniformity and several appendages extending from the polar
cap down to a latitude of $\sim 15^{\circ}$. Remarkably, neither a low-latitude band of spots nor 
an intermediate band free of spots
were found. The differences in the spot patterns found in the two investigations 
were explained by \citet{Barnesetal05} as a consequence 
of the higher value of the stellar $v \sin i$ adopted in the reconstruction made by \citet{Listeretal99}.
 As a matter of fact, \citet{Barnesetal05}
reconstructed a new map with their value of  $v \sin i$ using the line profiles by \citet{Listeretal99} and showed that the
resulting 1993 images were very similar to those obtained for the 1998 season, i.e., showing
 only high-latitude spots in addition to the polar cap. 
This result shows once again that the estimate of stellar parameters is a crucial task to obtain reliable Doppler maps of stars. 
{ In view of their accuracy and to warrant a better comparison with previous maps, we adopt the parameters of \citet{Barnesetal05}.} Our spectra
were acquired in  
September 2003, therefore they allow us to perform a relatively long-term study of the LO~Peg
 spot pattern in combination with those previous
studies. In particular, the presence of high-latitude spots appears to be a crucial feature to test stellar dynamo models
\citep[e.g., ][]{Bushby03,Covasetal05}. 

The surface differential rotation of LO~Peg was determined by \citet{Barnesetal05} by tracing the shear motion of starspots
along their sequence of observations that extended for seven days. They found an equatorial 
acceleration with an equator-pole lap time
of $181 \pm 35$ days, not too different from the Sun that has an equator-pole lap time of 120 days. 
However, the relative amplitude of
the surface differential rotation is about two orders of magnitude smaller than in the Sun,
 i.e., $\sim 0.002$, given the much shorter rotation period
of LO~Peg. Unfortunately, our sequence of observations spans only four nights, thus an independent determination of
the surface shear is not possible. 
On the other hand, the limited time interval of our observations with respect to the estimated equator-pole lap time makes differential rotation unrelevant for our analysis.

\section{Observations and data reduction}
\subsection{Observations}
\label{Observ}

LO~Peg  was observed on the nights between 12 and 15 September 2003 with
 the high-resolution spectrograph SARG\footnote{For details on SARG, see:
http://www.tng.iac.es/instruments/sarg/} at the 3.58-m {\it Telescopio Nazionale Galileo}, located 
at the Roque de Los Muchachos Observatory. SARG uses a R4 echelle grating (31.6 groves/mm) in a quasi-Littrow mode and has a
dioptric camera to image the cross-dispersed spectrum onto a mosaic of two $2048\times 4096$ EEV CCDs.

We collected 44 spectra in total, obtained with the yellow grism 
cross disperser (with a 300 grooves/mm echelle grating and a  blaze wavelength of 589 nm).
 Each spectrum consists of 54 extractable orders with a wavelength coverage ranging from 462  to 792 nm and has 
a resolution of $86,000$.
Table~\ref{table1} shows the journal of the observations for the four nights. 
During the third night, one spectra  was taken with a shorter time exposure, 
and after reduction it showed a too  low signal-to-noise ratio and was discarded.
For each night, flat fields and Ar-Th calibration lamps were also observed.
\begin{table*}
\caption{Journal of the observations of LO~Peg.}
\begin{tabular}{cccccc}
\hline
 & & & & & \\
Date              & UT start & UT end & Exp. time (s) & No. frames & S/N \\
 & & & & & \\
\hline 
& & & & & \\
12 September 2003 & 22:35    & 00:13   &  1800        &    4       &  50-80        \\
13 September 2003 & 20:55    & 04:06   &  1800        &  13        &  50-80        \\
14 September 2003 & 20:23    & 04:00   &  1800        &  12        & 50-80          \\
15 September 2003 & 20:02    &  04:18  &  1800        &  15       & 50-80   \\
& & &  & & \\
\hline
\end{tabular}
\label{table1}
\end{table*}

We did not observe any template star, contrary to \citet{Listeretal99}
 who observed, with the same telescope and spectrograph used for LO~Peg, 
the K5 dwarf Gl~673 (having an effective temperature similar to LO~Peg, but a significantly smaller $v \sin i$)
and 
the M1 dwarf Gl~649 
(with a temperature similar to that expected for the starspots on LO~Peg, i.e., $T_{\rm eff}\sim 3500$ K). 
{
For the reconstruction of the LSD line profiles, we adopted the LSD profile of Gl 673 as a template extracting its  spectrum from the Elodie archive}\footnote{See the online archive at http://atlas.obs-hp.fr/elodie/} (for details 
see Section~\ref{DItech}). 

Each spectrum of LO~Peg was obtained with an exposure time of 1800 s, giving a signal-to-noise 
ratio ranging from 50 to 80, at the 
cost of some phase smearing since the star rotates by $\sim 18^{\circ}$ during each exposure. 
This effect actually sets the surface
resolution of our Doppler maps, so that the use of the Elodie template, 
that has a resolution of $42,000$, does not introduce any
further  degradation of the map resolution. 
 
\subsection{Data Reduction}

Data reduction was performed using IRAF\footnote{IRAF is distributed by the National Optical Astronomy Observatories,
which are operated by the Association of Universities for Research
in Astronomy, Inc., under cooperative agreement with the National
Science Foundation.} (Image Reduction and Analysis Facility), separately for the blue and red CCDs. 
Rejection of hot and bad pixels is particularly important for our analysis as they can affect the LSD mean profiles \citep{DonatiCameron97}. 
Through a careful inspection, they were identified and replaced  
by performing a bi-linear interpolation 
along lines and columns using the nearest neighbour good pixels. 
We also discarded the wavelength range at the red end of each spectrum (above 6800\,\AA), badly affected by telluric lines.
Pixel-to-pixel gain variations were removed using flat-field lamp exposures. Images were then corrected for scattered light.

The echelle orders were extracted defining apertures on the best exposed LO~Peg frames. 
All the other spectra were extracted using the same frame apertures.
Wavelength calibration was finally performed using Th-Ar lamp spectra.

The shape of the continuum in the echelle orders was fitted by a $3^{\rm rd}$-order spline function.
To guide us in the selection of the continuum intervals for the normalisation,  
we broadened a synthetic spectrum with $T_{\rm eff}=4500$ K, $ \log g=4.5$ (cm s$^{-2}$), and ${\rm [Fe/H]}=0.5$ 
\citep{Coelhoetal05} with a rotational profile corresponding to the LO~Peg $v \sin i$ of  65.84 km s$^{-1}$
 \citep{Barnesetal05} and iteratively compared it with the continuum-normalised spectrum
 of LO~Peg until a satisfactory agreement was obtained. 

For phasing our data, we used the ephemeris of \citet{RobbCardinal95}:
\begin{equation}
\mbox {HJD of maxima } = 244 9909.8059 + 0.4236 \times E. 
\label{ephemeris}
\end{equation}

\section{Data Analysis}

\subsection{Least Square Deconvolution method}
\label{LSDdescr}

Doppler imaging of rapidly rotating stars ($v \sin i > 60-70$ km s$^{-1}$) is hampered by the 
shallow depths of the line profiles due to rotational broadening. This makes it difficult to measure the
distortion induced by photospheric spots and trace their motion along the line profiles as
the star rotates. A very high signal-to-noise ratio ($\geq 500-700$) is therefore needed to perform 
Doppler imaging on such objects and this is difficult to obtain, especially for a star as 
faint as LO Peg, because the exposure time is limited by the requirement that the  rotation of the
star during the acquisition of a spectrum be smaller than the resolution in longitude, in order to avoid 
a blurring of the resulting map. 

To solve this kind of problems, \citet{Donatietal97} and \citet{DonatiCameron97} introduced the Least Squares 
Deconvolution method (hereinafter LSD) that allows to extract a mean line profile with an enhanced signal-to-noise ratio 
from an entire spectrum by means of a suitable deconvolution technique. 
Its basic assumption is that the observed spectrum is given by the convolution of an unbroadened spectrum, represented in \citet{Donatietal97} and \citet{DonatiCameron97} by a series of Dirac delta functions, with  
a rotational broadening profile, assumed to have the same shape for all spectral lines.
Moreover, limb-darkening is assumed to be the same for all the spectral lines, i.e.,  independent of 
wavelength. These simplifying assumptions are not suitable for  spectral lines with an equivalent width 
larger than, say, $\sim 0.25$ \AA, because their profiles are significantly affected by the chromospheric temperature
increase and cannot be assumed to be similar to those of weaker lines. Therefore, the LSD technique can be applied only 
after removing spectral intervals containing chromospherically sensitive and strong spectral lines (see below for 
details).  

In the present work we used an improved version of the LSD technique, 
which substitutes a synthetic spectrum to a line-list in the deconvolution procedure. 
Using a synthetic spectrum instead of approximating the unbroadened spectrum 
as a sequence of Dirac delta functions, allows to take
into account the finite widths of the local line profiles resulting from, e.g.,  thermal Doppler broadening and
microturbulence. { Our approach is the same as that proposed by \citet{Barnes05} and we refer the reader to that
paper for more details}.

The rotational broadening profile contains the distortions induced by the surface brightness
inhomogeneities. By  convolution with the unbroadened spectrum, such distortions are reproduced along the
profile of each spectral line. In other words, the basic assumption of the technique is that all the spectral lines
repeat the same information on the rotational broadened profile. Therefore, it can be extracted by a suitable
deconvolution, when the unbroadened spectrum is known. The advantage of such an approach is that individual line profiles
are combined to enhance the signal-to-noise ratio of the extracted rotational broadened profile, thus allowing us to
perform Doppler imaging by using the sequence of such profiles. 

The mathematical formulation of the modified LSD method used in this paper is as follows. The flux in the spectrum 
at the wavelength $\lambda$, ${\cal F(\lambda)}$, 
normalized to the value of the corresponding continuum $F_{\rm c}(\lambda)$, can be written as:
\begin{equation}
{\cal F}(\lambda) \equiv \frac{F(\lambda)}{F_{\rm c}(\lambda)} = \int_{-v \sin i}^{v \sin i} S(\lambda -\lambda \frac{v_{\rm r}}{c}) {P}_{\rm rot} (v_{\rm r}) 
d v_{\rm r},
\label{line_broad}
\end{equation}
where $S(\lambda)$ is the normalized spectrum of a non-rotating star with the same atmospheric parameters, ${P}_{\rm rot}$ the
rotational broadened profile, $v_{\rm r}$ the radial velocity, ranging from $-v \sin i$ to $v \sin i$, with 
$v$ the equatorial velocity of rotation of the star, $i$ the inclination of its rotation axis, and $c$ the speed of light.
The rotational broadened profile is normalized so that: 
$\int_{-v \sin i}^{v \sin i} {P}_{\rm rot} (v_{\rm r}) d v_{\rm r} = 1$. 
Equation~(\ref{line_broad}) states that the observed normalized flux $ {\cal F}(\lambda)$
 is obtained by integrating the local, normalized specific intensity $S(\lambda)$, Doppler shifted at the radial velocity of the
considered point, over the stellar disk. Since the star is assumed to rotate rigidly,  the surface integration can be 
transformed into an integration over the radial velocity.  

If the brightness distribution over the photosphere of the
star is not homogeneous, specifically if dark spots are present, the rotational broadened profile can be 
written as:
\begin{equation}
{P}_{\rm rot} (v_{\rm r}) = w_{\rm u} {P}_{\rm u} (v_{\rm r}) + w_{\rm s} {P}_{\rm s} (v_{\rm r}),
\end{equation}
where ${P}_{\rm u}$ is the profile of the unperturbed photosphere and ${P}_{\rm s}$ that  of
 the spots, and the weights $w_{\rm u}$ and $w_{\rm s}$ are functions of the rotation phase that 
depend on $i$, $v \sin i$, the limb-darkening coefficient, the area and co-ordinates of the spots,
and the ratio of their specific intensity
to that of the unperturbed photosphere. 
{ To simplify our analysis, we assume that the spot line profile is simply a scaled version of 
the unperturbed  line profile, i.e., $P_{\rm s} = C_{\rm s} P_{\rm u}$, where $C_{\rm s}$ is a scaling constant whose value will be adjusted to warrant the best fit to the LSD line profiles.}

To transform the integration appearing in Equation~(\ref{line_broad}) into a discrete summation for computation
purposes, it is useful to consider the values of the local spectrum $S(\lambda)$ along a discrete set of wavelength
values, uniformly spaced in radial velocity, i.e.: $S_{i} \equiv S(\lambda_{i})$,
 where $\lambda_{i}= \left( 1 + \frac{\Delta v_{\rm r}}{c}\right) \lambda_{i-1}$, and  
$\Delta v_{\rm r}$ is the radial velocity interval the value of which is taken to sample adequately the unbroadened
line profiles. In the present analysis, we choose $\Delta v_{\rm r}=0.5$ km s$^{-1}$, i.e., significantly higher than the
resolution of the observed spectra because we use a unbroadened synthetic spectrum computed at a resolution of 
$200, 000$ by means of an LTE atmospheric model
with $T_{\rm eff} = 4500$ K, $\log g = 4.5$ (cm s$^{-2}$), microturbulence of 1.0 km s$^{-1}$ and abundance [Fe/H]$=0.5$ 
\citep[see][]{Coelhoetal05}.

We recast Equation~(\ref{line_broad}), expressing the integration 
with the extended trapezoidal rule:
\begin{equation}
{\cal F}_{i} = \sum_{k=0}^{N} a_{k} S_{i-k} {P}_{\rm rot \, \it k} \Delta v_{\rm r}, 
\label{eq_sys}
\end{equation}
where the index  $k$ specifies the radial velocity
according to: $v_{\rm r\; \it k} = -v \sin i + \Delta v_{\rm r} \times k$, $N = \frac{2 v \sin i}{\Delta v_{\rm r}}$, 
with $N+1$ being the number of points along the rotational broadened profile, and 
 the $a_{k}$'s are the numerical coefficients appearing in the
extended trapezoidal rule:
\begin{equation}
a_{k} = \left\{ 
\begin{array}{cl}
\frac{1}{2}  & \mbox{ for $k=0$}, \\
1  & \mbox{ for $0 < k \leq N-1$}, \\
\frac{1}{2} & \mbox{ for $k=N$}. \\
\end{array}
\right.
\end{equation}
The  rotational broadened profile ${P}_{\rm rot \; \it k}$ is specified by a set of $N+1$ values,
whereas the unbroadened spectrum consists of a much larger number of points $M$, as numbered by the
index $i$. Therefore, Equation~(\ref{eq_sys}) gives  a system of $M$ linear equations in $N+1$ unknowns, with
$M \gg N+1$, that is to be solved by the method of the least squares. Specifically, the solution
can be computed by means of the Singular Value Decomposition method (hereinafter SVD), 
as detailed in \citet{Pressetal92}. The elements of the
design matrix  for the problem at hand can be expressed as:
\begin{equation}
A_{ik} = \frac{a_{k} S_{i-k}}{\sigma_{i}},
\end{equation}
and the vector of the data as:
\begin{equation}
b_{i} = \frac{{\cal F}_{i}}{\sigma_{i}},
\end{equation}
where $\sigma_{i}$ is the standard deviation of the observed normalized flux ${\cal F}_{i}$ 
at wavelength $\lambda_{i}$. In addition to the least square solution of the system (\ref{eq_sys}), the SVD method
provides the covariance matrix among the $N+1$ values specifying the rotational broadened profile ${P}_{\rm rot; \it k}$, that
allows us to derive the errors on the mean profile and its effective resolution. This is particularly important in our case
because the resolution of the synthetic spectrum, used to specify the local profiles, is  higher than that of the
observed spectrum, so that neighbouring values of the solution vector ${P}_{\rm rot\; \it k}$ are expected to be correlated
(see Sect.~\ref{LSDprofiles}). { Moreover, \citet{Barnes05} noted that when the number of basis vectors adopted for 
the SVD solution is sufficiently high, high-frequency oscillations appear in the LSD profile owing to the fitting 
of residual noise. We performed a convolution of our LSD profiles with a Gaussian of FWHM of 4.0 km s$^{-1}$ to
overcome this problem, thus obtaining SVD2 profiles in Barnes' terminology. Given that our template spectrum has a resolution of 42,000, this treatment does not significantly degrade our map resolution.} 

Some wavelength intervals have been excluded from the SVD solution because they
include  strong lines or lines affected by the presence of the chromosphere. They are listed in Table~\ref{excluded_inter}.

The gain in the signal-to-noise ratio of the extracted mean broadened profile scales approximately 
as the square root of the number of spectral lines included into the LSD deconvolution. 
Considering about $600-800$ lines in the present application, 
the gain is $\sim 25$, that is we reach a final $S/N \simeq 2000$, on our best exposed spectra. 
\begin{table}
\caption{Wavelength intervals excluded from the Least Square Deconvolution. }
\begin{tabular}{cc}
\hline
 & \\
from & to \\
(nm) & (nm) \\
 & \\
\hline
 & \\
440.00  &  505.00  \\
514.50  &  519.00  \\
555.00  &  557.15  \\
574.00  &  578.20  \\
588.20  &  598.00  \\
612.50  &  632.00  \\
634.70  &  635.70  \\
654.90  &  657.80  \\
686.50  &  790.00  \\
 & \\
\hline
\end{tabular}
\label{excluded_inter}
\end{table}


\subsection{Doppler imaging technique}
\label{DItech}

The application of Doppler imaging to reconstruct a spot map from a sequence of mean broadened profiles, well distributed
in phase, requires local mean line profiles for the unperturbed photosphere and the spotted photosphere, respectively. We adopted the
LSD profile of the K5 dwarf Gl~673 ($T_{\rm eff} \sim 4500$ K) as the unperturbed profile and 
 assumed that the profile of the spotted photosphere is a simply scaled version of that of the
unperturbed photosphere, 
 as discussed in Sect.~\ref{LSDdescr}. The spectral resolution of the mean local profile 
 is $42, 000$.  
Note also that Gl~673 has a 
$v \sin i$ negligible in comparison to that of LO Peg, thus its mean profile can be considered unaffected by 
rotational broadening for our purposes. The widths of the instrumental profiles 
of SARG and Elodie are significantly smaller than the intrinsic widths of our mean profiles, 
so the fact that they have been
acquired with different spectrographs has a negligible effect on our map reconstruction. 

Synthetic mean profiles of LO~Peg were constructed for a trial distribution of the spot
 covering factor $f_{\rm s}$ on the surface of the star, 
according to the principles outlined in Sect.~\ref{LSDdescr}, where
$f_{\rm s}$ is the ratio of the area of the spots in a given surface element to that of the given element. 
To construct our maps, we adopted surface elements of size $9^{\circ} \times 9^{\circ}$ to oversample the actual resolution
elements that have an average size of $\sim 18^{\circ}$.  
 The parameters assumed to synthetize the mean 
rotational broadened profiles are: a) inclination of the rotation axis along the line 
of sight: $i = 45^{\circ}$; b) $v \sin i = 65.84$ km s$^{-1}$; c)  linear limb-darkening coefficient 
$u=0.8$; and d)  ratio of the intensity in the continuum of the spot mean profile 
to that of the unspotted mean profile $C_{\rm s}= 0.2$.  Inclination and $v \sin i$  were adopted from 
\citet{Barnesetal05}. We  show in Sect.~\ref{DIimages} that small variations of the values of 
$i$ and $v \sin i$ around the adopted values give  systematic residuals in the best fits of the line profiles 
thus confirming the accuracy of the parameters found by \citet{Barnesetal05}.
{Note that uncertainties in the limb-darkening coefficient and in the detailed shape of the local line profile do not significantly affect the reconstruction as they can be almost perfectly compensated by minute variations of the $v \sin i$ parameter at the level of $0.3-0.6$ km s$^{-1}$ \citep[cf., e.g., ][ for details]{UnruhCameron95}.
A variation of the contrast coefficient $C_{\rm s}$ corresponds to a change of the effective temperature of the spots and has an obvious impact on the covering factor of the spots. The adopted value corresponds to a spot temperature of $T_{\rm eff} \sim 3500$ K. }

Since LO~Peg is a fast rotator, we included the effects of  gravity 
darkening in the continuum adopting a stellar radius of 0.78 R$_{\odot}$, estimated from the $v \sin i$, the inclination and
the rotation period, and a mass of 0.8 M$_{\odot}$. The effective gravity was computed by adopting a simple Roche geometry
for the distorted star. 

As it is customary in Doppler imaging reconstruction, the trial spot distribution was varied to optimize an objective function
that consists of a linear combination of the $\chi^{2}$, obtained by fitting the sequence of the LSD profiles, and a
regularization function. Here we consider Maximum Entropy regularization (hereinafter ME)  according to the approach introduced by \citet{Cameron92}. 
A Lagrangian multiplier specifies the relative weight of the $\chi^{2}$ and of the ME  regularization in the
objective function. It is determined a posteriori by selecting the maximum value that leads to a distribution of the residuals
between the synthetic and the LSD profiles that does not deviate by more that one standard deviation from the 
Gaussian distribution obtained without any regularization. The algorithm of optimization is the same used by, e.g.,  
\citet[][]{Lanzaetal02}. { We extensively tested our code by simulating sequences of line profiles with added Gaussian noise for different spot configurations and reconstructing them through the procedure described above.} 

\section{Results}
\subsection{LSD profiles}
\label{LSDprofiles}
\begin{figure}
\centerline{\epsfig{file=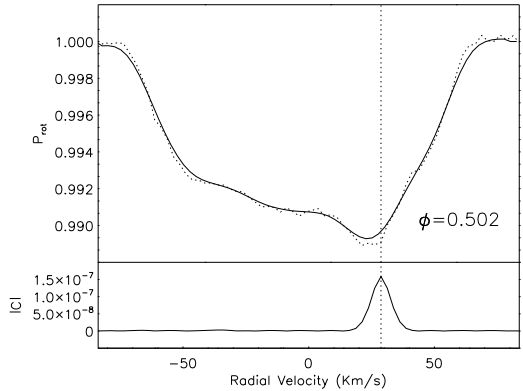,width=8cm}} 
\caption{{\it Upper panel:} Example of an LSD profile from a spectrum taken at rotation phase $\phi = 0.502$ on 13 September 2003. The flux, 
normalized at the continuum level, is plotted versus the radial velocity, measured from the centre of the LSD line profile.
The width of the radial velocity bins is 2.5 km s$^{-1}$.  The dotted line is the profile obtained from the SVD solution, while the solid line is the corresponding SVD2 profile (see the text). 
{\it Lower panel:} The covariance between the flux of the radial velocity bin marked by the vertically dotted line and those of
the adiacent bins along the LSD profile at the values of the radial velocity reported on the abscissa as computed with the SVD method before applying convolution with the Gaussian. }
\label{LSDsample}
\end{figure}

An example of the LSD profiles obtained with the procedure described in Sect.~\ref{LSDdescr} is plotted in the upper panel of 
Fig.~\ref{LSDsample}. { The dotted line is the LSD profile as obtained from the SVD solution, i.e., with  
small ripples due to residual noise fitting, while the solid line is the SVD2 profile obtained after a convolution with a Gaussian of FWHM of 4.0 km s$^{-1}$ \citep{Barnes05}.} 
The absolute value of the covariance between the flux value of one of the velocity bin  
along the profile (marked by a vertically dotted line) and those of its neighbour bins is plotted 
in the lower panel. The covariance is obtained with the SVD method before convolving with the Gaussian to obtain the SVD2 profile. Note that it is different from 
zero only in a narrow interval of amplitude $\sim 6$ km s$^{-1}$, centred around the given bin, 
 as estimated from the FWHM of the covariance curve.
 Similar values of the FWHM of the covariance curve are found also for the other velocity bins along the profile,
 so we conclude that the actual radial velocity resolution of our profiles is $\sim 6$ km s$^{-1}$. 
The maximum of the covariance curve, reached at the radial velocity value corresponding to 
the given bin, provides us with an estimate of the variance of the flux in that bin of the LSD profile. 
From the value in Fig.~\ref{LSDsample}, we find a
signal-to-noise ratio of $\sim 2000$ that confirms our a priori estimate in Sect.~\ref{LSDdescr}. 
 The signal-to-noise ratio is further increased by a factor of $\sim 2$ by the convolution with the Gaussian applied to obtain the SVD2 profiles. 

\begin{figure*}
\centerline{\epsfig{file=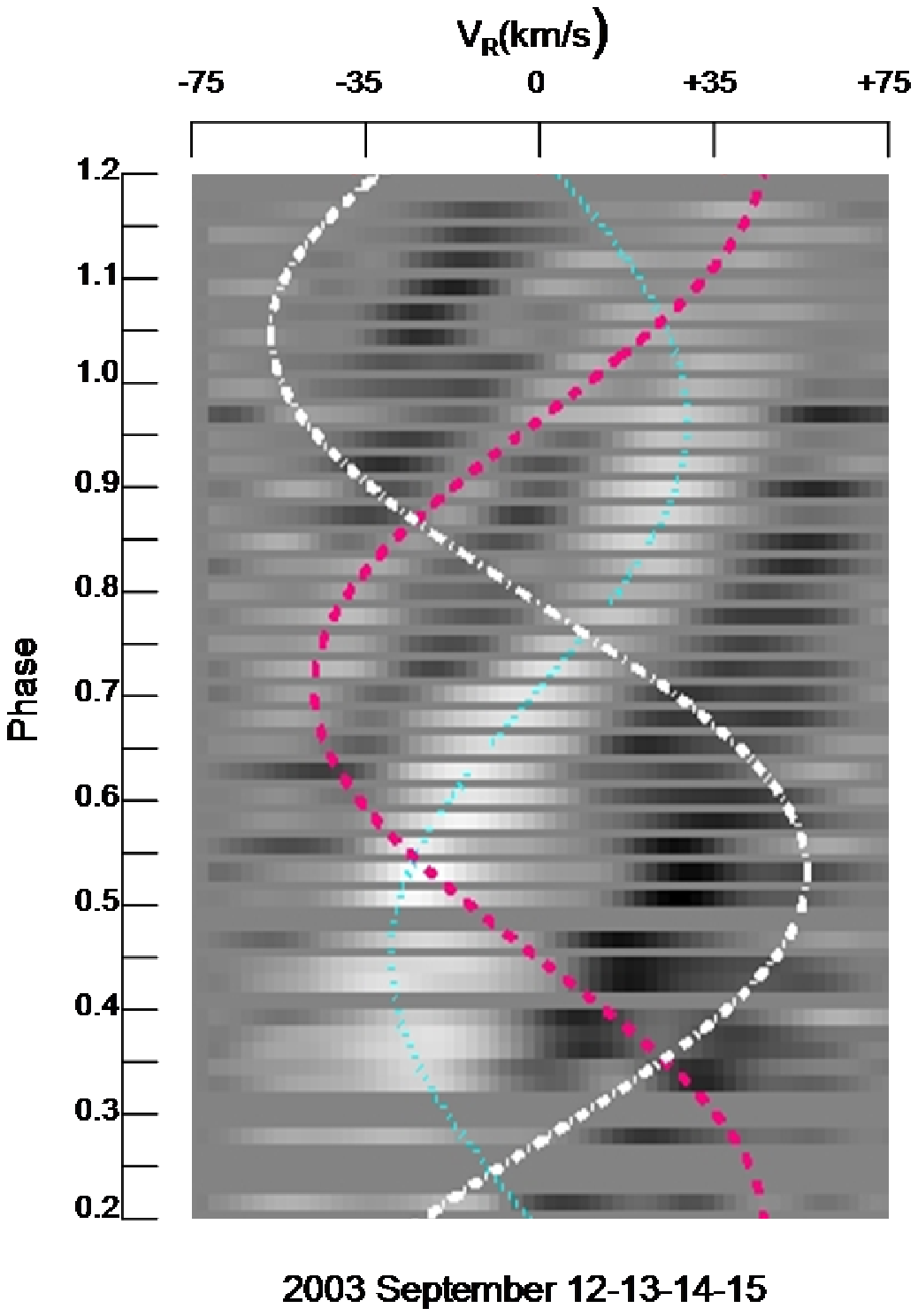,width=16cm}} 
\caption{Time series of the residual LSD line profiles, with rotation phase plotted versus radial velocity. The residuals are obtained by subtracting the grand average  profile from each profile.
White features correspond to starspot signatures (i.e.,  profile bumps). Three radial velocity curves have been traced by eye to help follow the signatures of the most prominent starspots. The spot parameters and the linestyles corresponding to each of them are listed in Table~\ref{spot_traces}, respectively.
}
\label{synoptic}
\end{figure*}

The time series of the residual LSD profiles, with rotation
phase plotted versus radial velocity, are shown in Fig.~\ref{synoptic}. { We put together all the profiles from  the four nights to have the best possible phase coverage}.  Our spectra were taken with an exposure of $\sim 1800$ s
that corresponds to their phase width on the plot. { The bumps corresponding to three individual starspots can be easily traced
along the sequence of the profiles. To help follow their motion, three curves have been superposed to the plot, each corresponding to one of the spots, respectively (see Table~\ref{spot_traces}). The phase at which a given bump crosses the central meridian of the star's disk depends 
on the longitude of the corresponding spot, while the amplitude of its radial velocity oscillation depends on the spot latitude. The fact that the radial velocity excursion is limited to approximately $\pm  (45-50)$ km s$^{-1}$ and the low inclination of the rotation axis of LO~Peg ($i = 45^{\circ}$) indicate that those starspots are
located at high latitudes ($\geq 40^{\circ}$).}
\begin{table}
\caption{Latitude $\varphi$ and phase $\phi$ of the transit across the central meridian of the stellar disk for the 
three spots corresponding to the traces in Fig.~\ref{synoptic}. The third column lists the linestyle and the colour of the corresponding trace in the same figure. }
\begin{tabular}{ccc}
\hline
 & & \\
$\varphi$ & $\phi$ & linestyle (color) \\
(deg) &  & \\
\hline
 & & \\
40 & 0.28 & dot-dashed (white) \\
50 & 0.95 & dashed (pink) \\
65 & 0.70 & dotted (light blue) \\ 
 & & \\ 
\hline
\end{tabular}
\label{spot_traces}
\end{table}

\subsection{Doppler image of LO~Peg}
\label{DIimages}

\begin{figure*}
\centerline{\epsfig{file=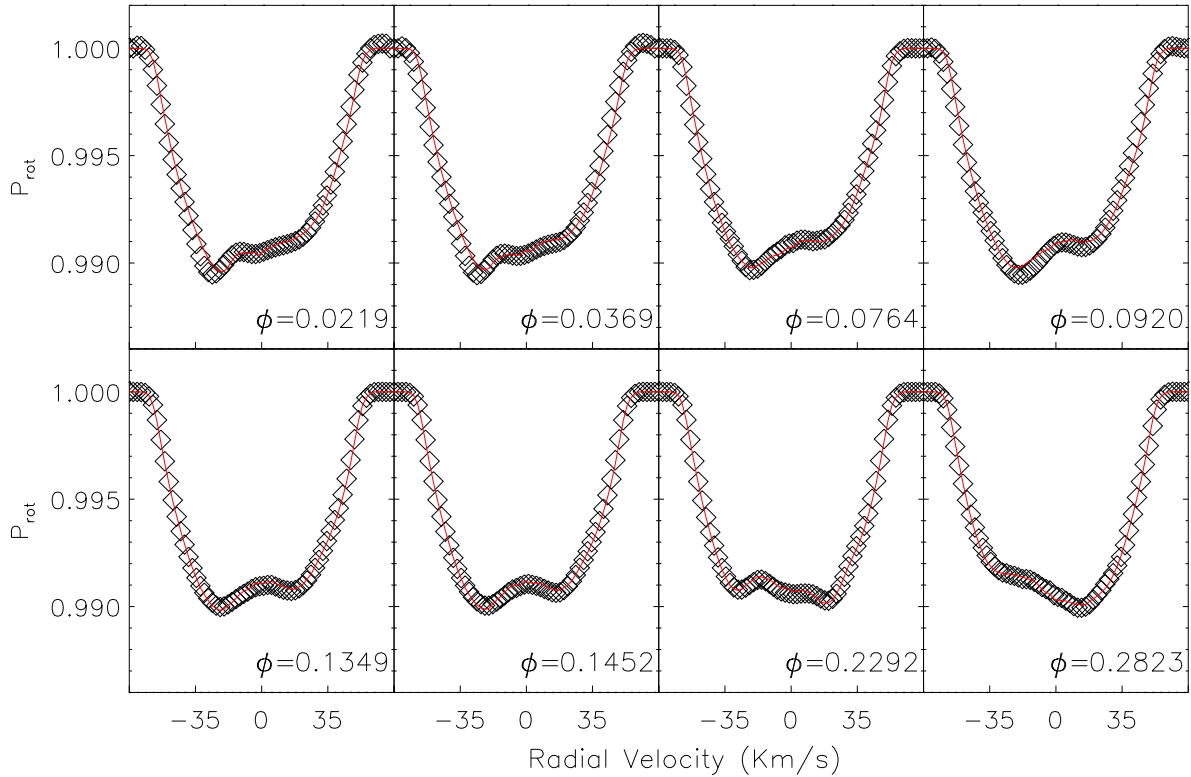,width=16cm}} 
\caption{A sequence of LSD profiles  (black open diamonds) 
with superposed their best fits computed with the ME regularization 
(red solid line), respectively. In each panel, the flux normalized
 to the continuum is plotted versus the radial velocity measured
 from the centre of the unperturbed profile. The rotation phase corresponding to each profile is reported in each panel and ranges from 0.0219 to 0.2823 for this subset of profiles. The dimensions of the plotting symbols correspond to one standard deviation of the
normalized relative flux.}
\label{profile_fits1}
\end{figure*}

\begin{figure*}
\centerline{\epsfig{file=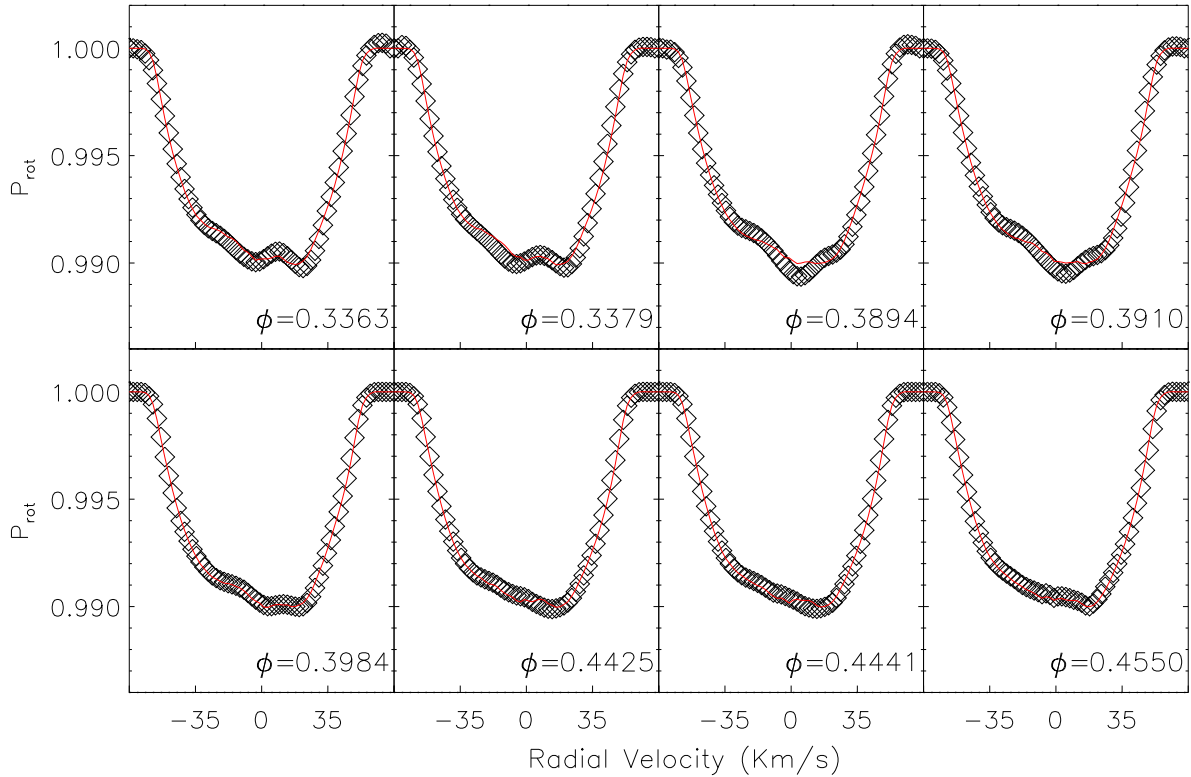,width=16cm}} 
\caption{As Fig.~\ref{profile_fits1} but for phases between 0.3363 to 0.4550.}
\label{profile_fits2}
\end{figure*}

\begin{figure*}
\centerline{\epsfig{file=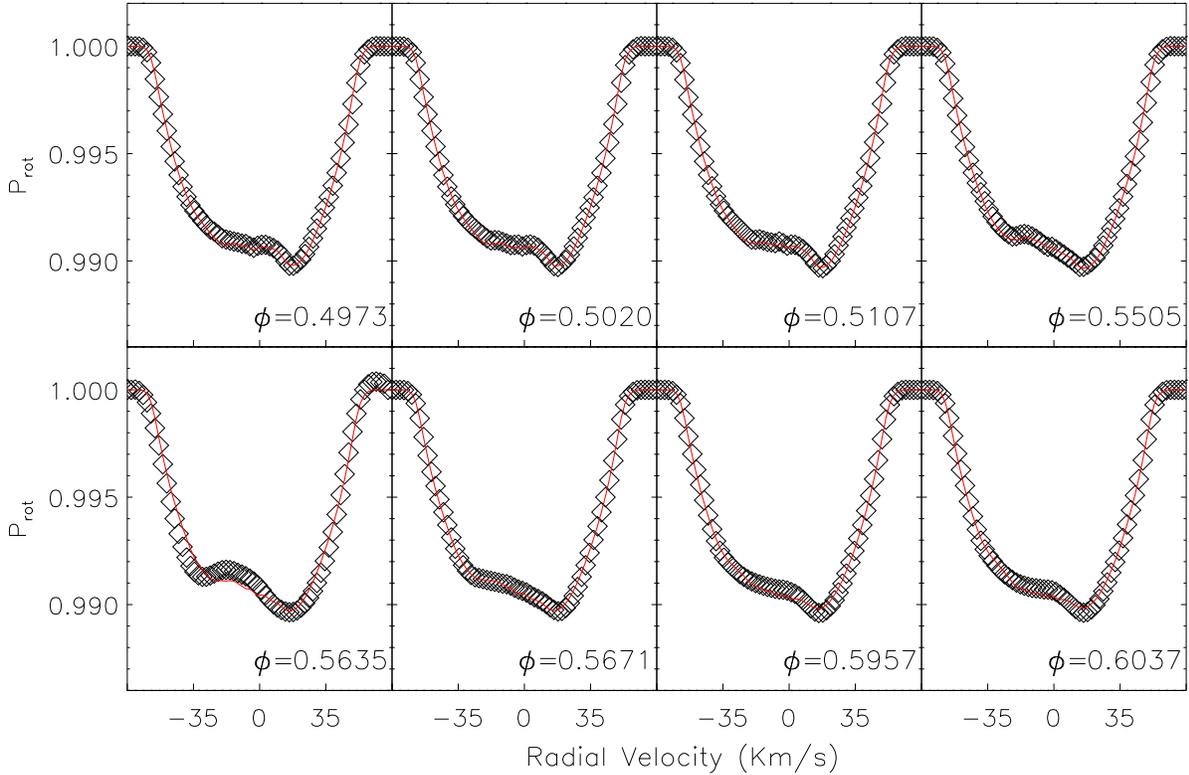,width=16cm}} 
\caption{As Fig.~\ref{profile_fits1} but for phases between 0.4963 to 0.6037.}
\label{profile_fits3}
\end{figure*}

\begin{figure*}
\centerline{\epsfig{file=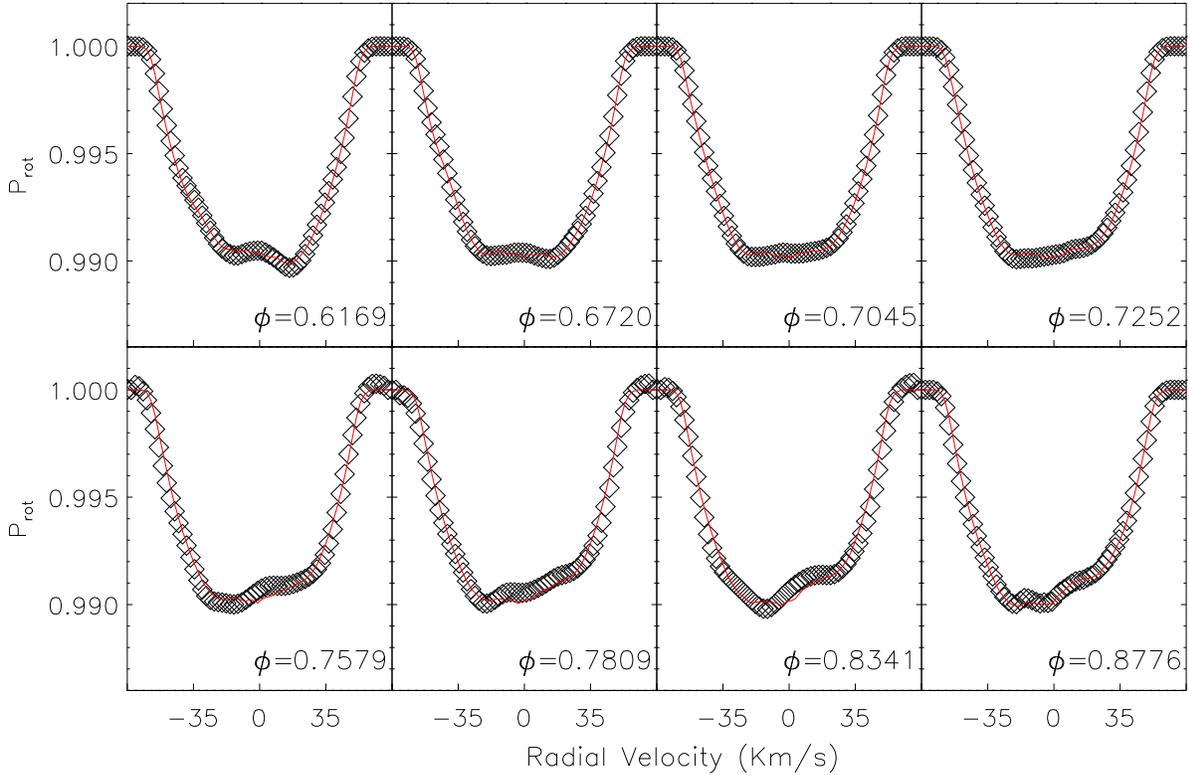,width=16cm}} 
\caption{As Fig.~\ref{profile_fits1} but for phases between 0.6169 to 0.8776.}
\label{profile_fits4}
\end{figure*}

\begin{figure*}
\centerline{\epsfig{file=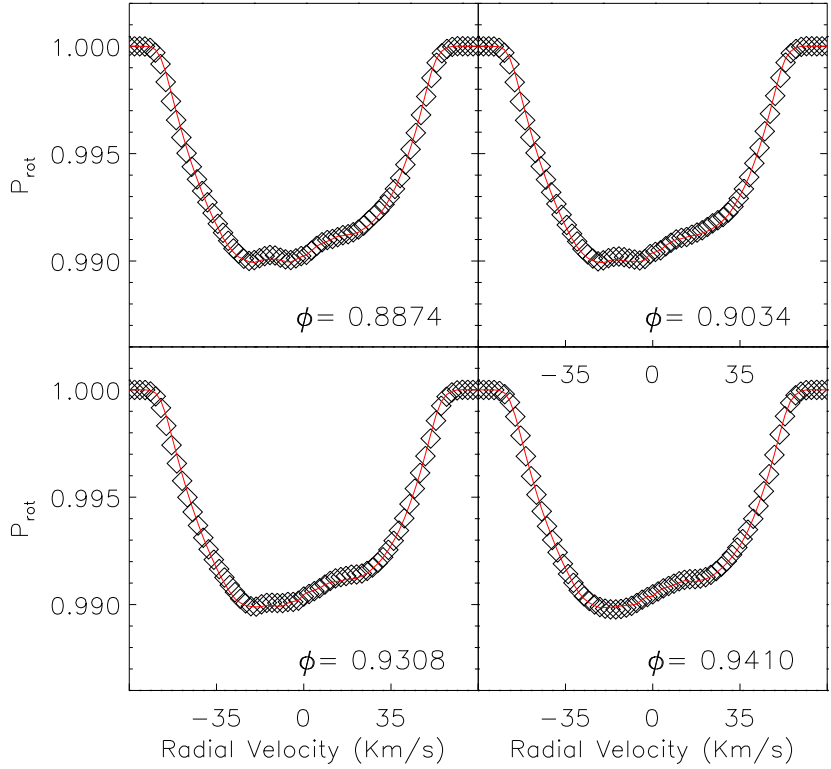,width=16cm}} 
\caption{As Fig.~\ref{profile_fits1} but for phases between 0.8874 to 0.9410.}
\label{profile_fits5}
\end{figure*}

The sequence of the best fits of the LSD profiles obtained with the ME regularization is shown 
in Figs.~\ref{profile_fits1}, \ref{profile_fits2}, \ref{profile_fits3}, \ref{profile_fits4} and \ref{profile_fits5}.
Only 38 profiles were actually fitted since 6 profiles coming from low-quality spectra were discarded. 
The fits are almost always very good, indicating that the spectral features associated with 
the mapped starspots are coherently traced as they move across the mean line profiles. 

\begin{figure}
\centerline{\epsfig{file=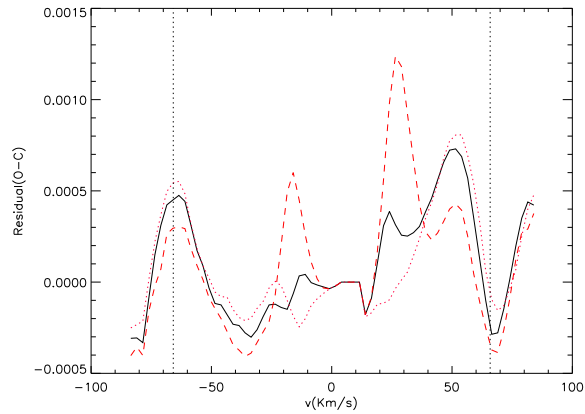,width=8cm}} 
\caption{Mean residual line profile for ME fits obtained for  $i=45^{\circ}$ \citep{Barnesetal05} (black solid line), $i=42^{\circ}$ (dotted red line) and $i=48^{\circ}$ (dashed red line). In all 
the cases the equatorial rotation velocity is fixed at $v = 93.11$ km s$^{-1}$, which corresponds to 
$v \sin i = 65.84$ km s$^{-1}$ for $i=45^{\circ}$.  }
\label{mean_prof_i}
\end{figure}

\begin{figure}
\centerline{\epsfig{file=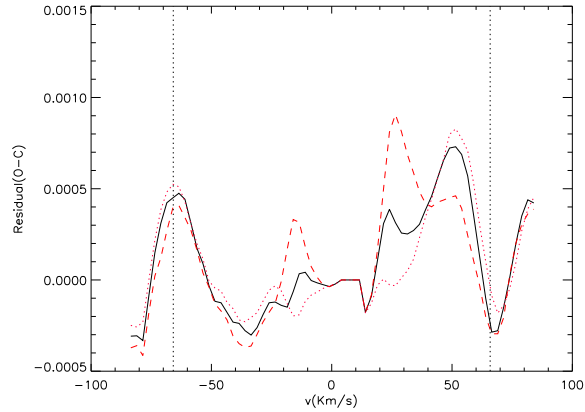,width=8cm}} 
\caption{Mean residual line profile for ME fits obtained for  $v \sin i=65.84$ km s$^{-1}$ \citep{Barnesetal05} (black solid line), $v \sin i= 63.72$ km s$^{-1}$ (dotted red line) and $v \sin i= 67.96$ km s$^{-1}$ (dashed red line). In all the cases $i = 45^{\circ}$. }
\label{mean_prof_veq}
\end{figure}

To show the effect of a small variation of the inclination $i$ and $v \sin i$ on the ME best fits to the 
line profiles, we plot in Figs.~\ref{mean_prof_i} and \ref{mean_prof_veq} the mean residual line 
profiles for different values of those parameters. The values given by \citet{Barnesetal05} are those giving 
the best average fit while different values produce larger residuals in some 
radial velocity intervals \citep[cf. Fig.~3 of ][]{Barnesetal05}.

\begin{figure*}
\centerline{\epsfig{file=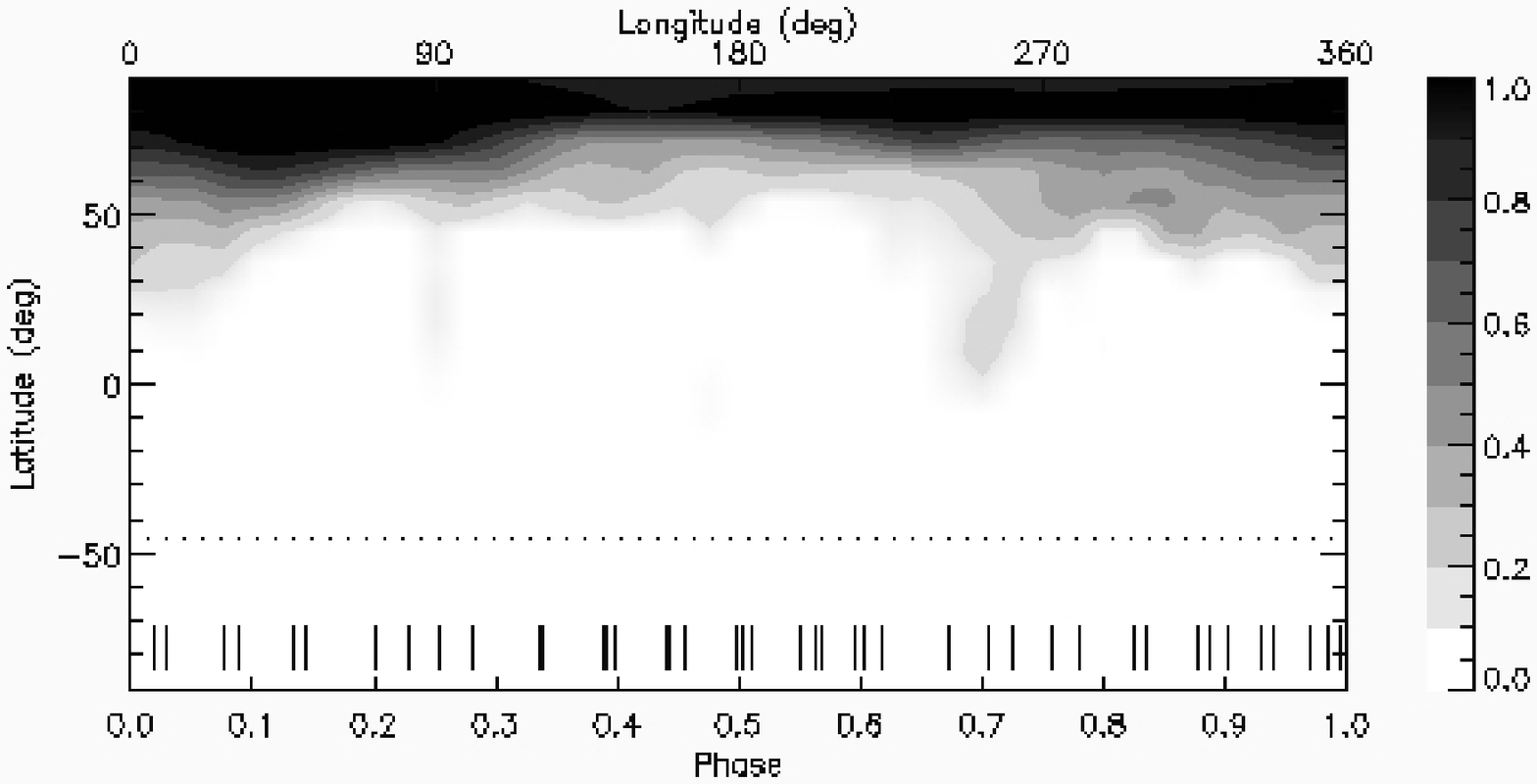,width=16cm}} 
\caption{The ME-regularized map of LO~Peg for 12-15 September 2003. The horizontal dotted line marks the border between the
visible and  invisible parts of the star (the adopted inclination of the stellar rotation axis is $45^{\circ}$).
The small vertical ticks close to the bottom of the panel mark the rotation phases at which the observations were made. 
 The scale on the
right indicates the grey level corresponding to the spot covering factor $f_{s}$.}
\label{MEmap}
\end{figure*}

In order to improve the phase coverage, the LSD profiles of
all the four nights from 12 to 15  September  were combined together to derive a single ME-regularized 
map of the surface of LO~Peg, shown in Fig.~\ref{MEmap}. 
{ The data are fitted at a reduced $\chi^{2}$ level of 1.1, for a mean standard deviation of $\sigma=2.1 \times 10^{-4}$ of the continuum flux obtained for the SVD2 profiles. } 
The longitude resolution of the map is  limited by the phase blurring due to the exposure time and is
$\sim 18^{\circ}$. The latitude resolution is a function of the latitude itself, with an 
average value of $\sim 20^{\circ}$ at a latitude of $\sim 50^{\circ}$, as derived from our radial velocity resolution of $\sim 6$ km s$^{-1}$ (cf. Sect.~\ref{LSDprofiles}). 

Our ME map is similar to those of \citet{Barnesetal05}, showing a concentration of starspots at latitudes
higher than $60^{\circ}$ with a covering factor varying with longitude. Appendages are also found, extending down to
a latitude of $10^{\circ}-20^{\circ}$ and corresponding to the moving bumps we traced on the line profile sequence in Fig.~\ref{synoptic}. { Those features are the most reliable and their reconstructions are largely insensitive to  uncertainties in the mapping parameters (cf. Sect.~\ref{DItech}) because they can be directly traced onto the line profile sequence.} 

In comparison with the maps of \citet{Barnesetal05},
our map  indicates a slightly higher level of mid- and 
low-latitude activity. However, this result
should be taken with caution because the latitude resolution of the Doppler imaging technique is lower close to the equator. 
Moreover,  ME regularization sometimes tends to produce small concentrations of spots to improve the fit of the line profiles,
especially at phases where observations are missing \citep[the so-called {\it superresolution effect}, see, e. g., ][]{NN86}. { The small spot just at the threshold of visibility close to the equator at a longitude of $\sim 170^{\circ}$ may therefore be an artefact as suggested by the lack of an obvious corresponding bump in the sequence of line profiles and the fact that it quickly disappears when the Lagragian parameter is increased.} 


\section{Discussion and conclusions}

We used the LSD approach of \citet{Barnes05}, based on  a synthetic spectrum instead of
a line list, as adopted in the original formulation of \citet{DonatiCameron97}. 
Our approach takes into
account the finite width of the spectral lines due to, e.g.,
the thermal Doppler, pressure and microturbulence broadenings, allowing us a 
very good fit of the observed spectra. The advantage 
is especially apparent  in the match of our LSD  
profiles to the  continuum in the red and blue extreme line wings. 
Moreover, the use of the covariance matrix obtained through the SVD technique, allows us 
an a posteriori check of the signal-to-noise ratio along the average profiles 
and a determination of their actual radial velocity resolution.

Our Doppler images of LO~Peg are similar to those of \citet{Barnesetal05}, with a polar
cap and some starspot appendages extending down to $\approx 10^{\circ}$ latitude. 
The spot concentrations at high latitudes are largely independent of our Doppler imaging assumptions,
as the moving features associated with them are clearly visible on the time sequence of residual line profiles 
in Fig.~\ref{synoptic}. 
The polar cap is not symmetric around the pole, being more extended around phase 0.1 and
less extended around phase $0.55$. 

Our maps, as well as those of \citet{Barnesetal05}, do not show starspots southern than $\sim 10^{\circ}$.
However, the presence of small spots or of a band of uniformly distributed spots at low latitudes cannot be ruled out
on the basis of our Doppler image. The former may escape detection due to foreshortening effects, since the 
inclination of the stellar rotation axis is quite low. The latter does not produce any migrating feature on the
average line profile, so any systematic error in the local line profile or stellar parameters can 
effectively prevent its detection.

The long-term persistence of the polar spot on LO~Peg is particularly intriguing when its behaviour is
compared with those of other rapidly rotating late-type stars \citep[see Sect. 4.1 of ][]{Barnesetal05}.
The possibility that the polar spot is a non-persistent feature is supported by the Doppler maps 
of AB~Dor \citep[][]{Kursteretal94} that are more numerous and better distributed in time than those of LO~Peg. 
The predominance of the Coriolis force on magnetic buoyancy and magnetic tension forces on
flux tubes emerging from the base of the convection zone or the overshoot region at the interface with the
radiative core, can explain the appearance of flux at intermediate and high latitudes in rapidly rotating stars
\citep[see, e.g, ][]{Schussleretal96}.
In combination with surface diffusion and poleward meridional flows, such models can explain the
presence of polar spots and their long lifetimes \citep[e.g., ][]{SchrijverTitle01,Isiketal07}.
The concentration of the magnetic flux toward the poles has relevant effects also on the 
angular momentum and mass loss of such stars through their magnetized winds \citep[see, e.g., ][]{Buzasi97,Holzwarth05}.

The presence of low-latitude spots is difficult to explain on the basis of a model considering
slender flux-tubes originating in the overshoot layer or near the base of the convection zone and points
toward the existence of a distributed turbulent dynamo working throughout most of the convection zone, as in the
models considered by, e.g., \citet{Bushby03} and \citet{Covasetal05}. Further support for the presence of a mean-field
distributed dynamo in the convection zones of late-type rapidly rotating stars comes from the 
works of \citet{Donati99} and \citet{Lanza06,Lanza07}.

\section*{Acknowledgments}

The authors are grateful to an anonymous Referee for a careful reading of the manuscript and valuable comments. 
This work is based on observations made at the {\it Telescopio Nazionale Galileo} of the 
{\it Istituto Nazionale di Astrofisica} (INAF) whose staff is gratefully acknowledged for 
their kind assistance during the operation and the acquisition of data with the SARG spectrograph.
The authors wish to thank also Drs. Salvo Scuderi, Paolo Romano, Giuseppe Cutispoto, Antonio Frasca, Katia Biazzo
 and Marco Comparato for  interesting discussions.  
Active star research at INAF-Catania Astrophysical Observatory and the Department of Physics
and Astronomy of Catania University is funded by MUR ({\it Ministero dell'Universit\`a e della Ricerca}), 
and by {\it Regione Siciliana}, whose financial support is gratefully
acknowledged.

This research has made use of the ADS-CDS databases, operated at the CDS, Strasbourg, France.


\begin{thebibliography}{99}

\bibitem[\protect\citeauthoryear{Barnes}{2004}]{Barnes05}
Barnes, J. R., 2004, MNRAS, 348, 1295 

\bibitem[\protect\citeauthoryear{Barnes et al.}{2005}]{Barnesetal05}
Barnes, J. R., Collier Cameron, A., Lister, T. A., Pointer, G. R., Still, M. D. 2005,
MNRAS, 356, 1501 

\bibitem[\protect\citeauthoryear{Bushby}{2003}]{Bushby03}
Bushby, P. J. 2003, MNRAS, 342, L15

\bibitem[\protect\citeauthoryear{Buzasi}{1997}]{Buzasi97}
Buzasi, D. L., 1997, ApJ, 484, 855

\bibitem[\protect\citeauthoryear{Coelho et al.}{2005}]{Coelhoetal05}
Coelho, P., Barbuy, B., Melendez, J., Schiavon, R., Castilho, B., 2005, A\&A 443, 735

\bibitem[\protect\citeauthoryear{Collier Cameron}{1992}]{Cameron92}
Collier Cameron, A. 1992, LNP, 397, 33 

\bibitem[\protect\citeauthoryear{Covas et al.}{2005}]{Covasetal05}
Covas, E., Moss, D., Tavakol, R. 2005, A\&A, 429, 657

\bibitem[\protect\citeauthoryear{Donati}{1999}]{Donati99}
Donati, J.-F. 1999, MNRAS, 302, 457 

\bibitem[\protect\citeauthoryear{Donati \& Collier Cameron}{1997}]{DonatiCameron97} 
Donati, J.-F., Collier Cameron A. 1997, MNRAS, 291, 1

\bibitem[\protect\citeauthoryear{Donati et al.}{1997}]{Donatietal97} 
Donati, J.-F., Semel, M., Carter, D. M., Rees, D. E., Cameron, A. C. 1997, MNRAS 291, 658

\bibitem[\protect\citeauthoryear{Holzwarth}{2005}]{Holzwarth05}
Holzwarth, V. 2005, A\&A, 440, 411

\bibitem[\protect\citeauthoryear{I\c{s}{\i}k et al.}{2007}]{Isiketal07}
I\c{s}{\i}k, E., Sch\"ussler, M., Solanki, S. K. 2007, A\&A, 464, 1049

\bibitem[\protect\citeauthoryear{Jeffries et al.}{1994}]{Jeffersetal94}
Jeffries, R. D., Byrne, P. B., Doyle, J. G., Anders, G. J., James, D. J., Lanzafame, A. C. 1994,
MNRAS, 270. 153 

\bibitem[\protect\citeauthoryear{K\"urster et al.}{1994}]{Kursteretal94}
K\"urster, M., Schmitt, J. H. M. M., Cutispoto, G. 1994, A\&A, 289, 899

\bibitem[\protect\citeauthoryear{Lanza}{2006}]{Lanza06}
Lanza, A. F. 2006, MNRAS, 373, 819

\bibitem[\protect\citeauthoryear{Lanza}{2007}]{Lanza07}
Lanza, A. F. 2007, A\&A, 471, 1011 

\bibitem[\protect\citeauthoryear{Lanza et al.}{2002}]{Lanzaetal02}
Lanza, A. F., Catalano, S., Rodon\`o, M., $\dot{\rm I}$banoglu, C., Evren, S., et al. 2002, A\&A, 386, 583 

\bibitem[\protect\citeauthoryear{Lister et al.}{1999}]{Listeretal99}
Lister, T. A., Collier Cameron, A., Bartus, J. 1999, MNRAS, 307, 685  

\bibitem[\protect\citeauthoryear{Narayan \& Nityananda}{1986}]{NN86}
Narayan R., Nityananda R., 1986, ARA\&A 24, 127

\bibitem[\protect\citeauthoryear{Piskunov et al.}{1990}]{Piskunovetal90}
Piskunov, N. E., Tuominen, I., Vilhu, O. 1990, A\&A, 230, 363 

\bibitem[\protect\citeauthoryear{Press et al.}{1992}]{Pressetal92} 
Press, W. H., Teukolsky, S. A., Vetterling, W. T., Flannery, B. P. 1992, Numerical Recipes,
2nd edn., Cambridge Univ. Press, Cambridge 

\bibitem[\protect\citeauthoryear{Robb \& Cardinal}{1995}]{RobbCardinal95}
Robb, R. M., Cardinal, R. D. 1995, IBVS, 4221, 1  

\bibitem[\protect\citeauthoryear{Sch\"ussler et al.}{1996}]{Schussleretal96}
Sch\"ussler, M., Caligari, P., Ferriz-Mas, A., Solanki, S. K., Stix, M. 1996, A\&A, 312, 503

\bibitem[\protect\citeauthoryear{Schrijver \& Title}{2001}]{SchrijverTitle01}
Schrijver, C. J., Title, A. M., 2001, ApJ, 551, 1099

\bibitem[\protect\citeauthoryear{Unruh \& Collier Cameron}{1995}]{UnruhCameron95}
Unruh, Y. C., Collier Cameron A., 1995, MNRAS, 273, 1

\end{thebibliography}
\end{document}